\documentclass{aa}
\usepackage[varg]{txfonts}
\usepackage{graphicx}
\usepackage{amsmath}
\usepackage{hyperref}
\usepackage{multirow}
\usepackage{lscape}
\usepackage{longtable}
\usepackage{xcolor}

\usepackage{hyperref}

\newcommand{\kmsmpc}{km s$^{-1}$Mpc$^{-1}$}

\newcommand\hmpc{$h^{-1}$Mpc}
\newcommand{\kms}{km s$^{-1}$}

\newcommand{\h}{$h^{-1}$}

\DeclareUnicodeCharacter{2003}{}
\DeclareUnicodeCharacter{2212}{}

\begin{document}

\title{In search for the Local Universe dynamical homogeneity scale with CF4++ peculiar velocities}

\author{H.M.~Courtois\thanks{h.courtois@ip2i.in2p3.fr}\inst{1} ,
J.~Mould\inst{2,3},
A.M.~Hollinger\inst{1},
  A.~Dupuy\inst{4},
 C.P.~Zhang\inst{5}
}

\institute{
Universit\'e Claude Bernard Lyon 1, IUF, IP2I Lyon, 4 rue Enrico Fermi, 69622 Villeurbanne, France
\and
Centre for Astrophysics \& Supercomputing, Swinburne University, Hawthorn, VIC 3122, Australia
\and
ARC Centre of Excellence for Dark Matter Particle Physics
\and
Korea Institute for Advanced Study, 85, Hoegi-ro, Dongdaemun-gu, Seoul 02455, Republic of Korea
\and
National Astronomical Observatories, Chinese Academy of Sciences, Beijing 100101, China
} 

\date{Received A\&A 2024; Accepted date}

\abstract {
This article explores an update to the cosmography of the local Universe within $z=0.1$, incorporating galaxy peculiar velocity datasets from the first data releases of WALLABY, FAST,  and DESI surveys. The galaxies with peculiar velocities currently selected in each survey is 655, 4796, and 4191 respectively. The new CF4++ compendium enables a more comprehensive study of the nearby Universe bulk flow dynamics. We find a bulk flow of $315 \pm 40$ \kms\ at 150 \hmpc. 
This analysis additionally reveals that the dynamical scale of homogeneity is not yet reached in the interval [200-300] $h^{-1}$Mpc from the observer. 
This new data also refines the structure of local superclusters, revealing more spherical shapes and more clearly defined boundaries for key regions such as Great Attractor (Laniakea) and Coma. Very few measurements make a big difference in revealing the hidden Vela supercluster.
To help colleagues obtain a peculiar velocity prediction for the object they are currently studying (SNIa, gravitational waves, etc.), we publicly release the CF4++ catalogue, as well as the reconstructed density and velocity fields used in this work.}


\titlerunning{Dynamical homogeneity scale}
\authorrunning{Courtois et al.} 
\maketitle

%
%-------------------------------------------------------------------

\section{Introduction}

Near-field cosmology explores the structure, dynamics, and evolution of the Universe on relatively small scales within the redshifts $ z < 0.1 $. A crucial tool in this domain is the analysis of peculiar velocity datasets, which capture the deviations of galaxies' motions from the Hubble flow due to gravitational interactions. By studying these peculiar velocities, we can infer the distribution of mass, both visible and dark, within the local cosmic volume. This allows for detailed examinations of the density field, the local matter power spectrum, and tests of cosmological models and alternative theories of gravity. Peculiar velocity measurements also complement large-scale observations, such as cosmic microwave background and redshift surveys, by constraining parameters like the growth rate of structure formation and the amplitude of matter clustering $\sigma_8 $. 

The bulk flow as a cosmological probe refers to the coherent motion of galaxies within a specific volume of the universe, driven by gravitational interactions with the large-scale distribution of matter. In the local Universe, this motion represents a deviation from the expected uniform expansion of the Hubble flow, providing a unique diagnostic of the matter distribution, including dark matter, on scales of tens to hundreds of megaparsecs.  

The study of the bulk flow in the local Universe provides insights into the scale of gravitational influences, the validity of the standard $\Lambda$CDM cosmological model, and potential deviations such as anisotropies or departures from Gaussianity in the primordial density field. 

Measurements of the bulk flow are typically derived from peculiar velocity datasets, which quantify the velocity of galaxies relative to the smooth expansion of the universe. These velocities are obtained through various observational techniques, including distance-independent indicators like the Tully-Fisher relation or the fundamental plane \citep{N11,H14,S16,Q18,Q21,A23}, and supernovae Ia luminosities \citep{C11,D11,T12,B20,L24}, a combination of several distance indicators in composite catalogues \citep{W09,H15,Q19,B20,C23,W23,WH23,B24} or redshift surveys \citep{C15,S21}.

This article presents an updated cosmography of the local Universe within $z=0.1$ using the composite catalogue Cosmic-Flows-4 originally consisting of 55,876 galaxies and upgraded by the addition of 9,642 galaxies from the first data releases of the WALLABY, FAST, and DESI surveys, called CF4++. We first present the building of the CF4++ dataset in section~\ref{sec:datasets}, and the updated local Universe density and gravitational velocity fields reconstruction in section~\ref{sec:reconstruction}.  Section~\ref{sec:results} delivers the analysis of the bulk flow in the Local Universe and conclusions are drawn in section~ \ref{sec:conclusion}.

\section {Integration of recent peculiar velocity datasets}\label{sec:datasets}
\subsection{WALLABY Pilot Survey}
The Widefield ASKAP L-band Legacy All-sky Blind surveY  \citep[WALLABY;][]{2020Ap&SS.365..118K} SKA pilot survey recently released its first set of public data, marking a significant milestone for the project (\cite{2022PASA...39...58W},
\cite{2024PASA...41...88M}). Conducted using the Australian Square Kilometre Array Pathfinder (ASKAP), WALLABY aims to map neutral hydrogen (HI) across the Southern Hemisphere. This data release includes HI spectral line cubes and catalogues from three pilot survey regions, providing detailed information on galaxy distributions, dynamics, and gas content.  

These observations offer valuable insights into galaxy evolution, large-scale structure, and cosmic flows. The release also serves as a test-bed for data processing and analysis pipelines in preparation for larger SKA-scale surveys. Researchers worldwide can now explore this publicly available dataset to conduct their own studies and collaborate on advancing our understanding of the Universe.
In  \cite{2023MNRAS.519.4589C} and \cite{2024MNRAS.533..925M}, we have shown how to transform the WALLABY pilot survey data into distance moduli that can be later interpreted as peculiar velocities. In this article we use 655 WALLABY galaxy peculiar velocities.

\subsection{FAST DR1 Dataset}

The 2024 data release from the Five-hundred-meter Aperture Spherical Telescope \citep[FAST;][]{2024SCPMA..6719511Z} represents a significant milestone in radio astronomy. This release, part of the FAST All-Sky HI Survey (FASHI), includes a catalogue of over 41,700 extragalactic neutral hydrogen (HI) sources, making it one of the largest HI surveys on a single telescope dataset. With a sensitivity and resolution superior to previous surveys, such as ALFALFA, the data enable detailed studies of galaxy dynamics, star formation, and the large-scale structure of the Universe. The survey also provides crucial cross-matched catalogues linking FAST data to optical sources, further enhancing its scientific utility.

This publicly available dataset not only supports cosmological research but also sets the stage for groundbreaking studies in galaxy formation and evolution, as well as follow-ups using FAST's advanced capabilities. By releasing this unprecedented dataset, FAST solidifies its role as a leading facility in global radio astronomy.

\cite{2024SCPMA..6719511Z} have provided optical identification for 10,976 of their sources with the Siena Galaxy Atlas \citep[SGA;][]{2023ApJS..269....3M}. From the $grz$ photometry of the SGA we chose the $z$ band for the Tully-Fisher relation (TFR), because of its lesser problems with internal galactic extinction. The SGA authors used profile fitting to obtain total magnitudes for their galaxies, simultaneously solving for ellipticity, and cataloguing axial ratios, which are necessary for the TFR. 
\cite{2024SCPMA..6719511Z} have compared their detection velocities with those of the ALFALFA survey \citep{2018ApJ...861...49H}, finding good agreement. We compared their W$_{50}$ values with those of ALFALFA and also found good agreement using 1938 galaxies. 

For measuring distances, CF4 uses the baryonic TFR. In order to incorporate the FAST galaxies, we used the $g-r$ colour to derive
the galaxy's $M_*/L_z$. We modify equation (1) of \cite{2015MNRAS.446.2144T} using the following equations,
\begin{equation}
    g-i = 1.408 (g-r -0.753)  + 1.14 
\end{equation}
and 
\begin{equation}
    i-z = 0.03 (g-r -0.753) + 0.235 \\,
\end{equation}
 obtained from tight colour-colour relations fitted to $ugriz$ photometry of Hydra galaxies by \cite{2021MNRAS.500.1323L}. Which gives the adopted relation 
 \begin{equation}
     \log M_* = 0.974 (g-r) + 0.802 -0.4 M_z \\.
 \end{equation}
 
 We normalized the obtained $M_*$ values to those of 850 galaxies in the FAST-SGA sample that also appear in the CF4 catalogue. This normalization has an uncertainty of 0.0046 dex. We similarly normalized the FAST HI masses to the gas masses of the CF4 catalogue with an uncertainty of 0.018 dex.  The combined uncertainty of the zero-point of the baryonic TFR is closer to the $M_*$ zero-point uncertainty than to the gas mass  uncertainty, as galaxies on average have a gas mass which is less than their stellar mass. This normalization also brings the data into conformity with the Hubble Constant inferred from the CF4 catalogue of 77.2 \kmsmpc.

%\textcolor[red]{
We omit galaxies whose magnitudes exhibit disagreements greater than 2.5$\sigma$ between FAST/Siena and CF4 from the sample. We additionally omit any objects with fractional FAST W$_{50}$ errors exceeding  4.7\%.
The $z$ band magnitudes were corrected for extinction using the Caltech-IPAC E(B-V) values for each galaxy individually \citep{2011ApJ...737..103S}. This results in minor corrections, as the biggest E(B-V) was 0.2 and the median $<$ 0.02 mag, with the  ratio of the total-to-selective extinction A$_z$/E(B-V) = 1.263. Finally, some galaxies were removed from the distance measurement sample as 2.5$\sigma$ deviants from the TFR fit.
%}

\subsubsection{Analysis of galaxies identified with Sloan Digital Sky Survey galaxies}
Some 14,070 FAST galaxies are identified with SDSS objects. We cross-matched these objects with DELVE DR2 objects 
from the DECam Local Volume Exploration Survey \citep[DELVE DR2 catalogue;][]{2022ApJS..261...38D})
with a matching radius of 7$^"$. Based on images with the Dark Energy Camera at Cerro Tololo in Chile, DELVE has limited coverage of the Northern Hemisphere.  We use the observed 21 cm velocity width data, $\Delta V$, from FAST and correct for the galaxies' inclination angle, $i$,  derived from the SGA axial ratios, 
to obtain a measurement  of $\Delta V(0) = \Delta V/\sin i $. 
Figure \ref{fig:TFR} shows the baryonic TFR of the galaxies incorporated into CF4++. A least-square fit yields the relation $\log_{10}M_\text{baryonic}=10.27+2.65(\Delta V(0)-2.36)$.  Galaxies that demonstrate deviations greater than 2.5$\sigma$ from this fit were discarded from the distance sample.  %deviations  

The baryonic TFR normalization of $M_*$ was done using the full sample of DELVE DR2 galaxies that are in the CF4 catalogue, only one of which is a FAST detection. The zero-point of this relation is determined by 54 galaxies to 0.06 dex in $\log_{10} M_*$.  The test for zero-point accuracy is whether the value of H$_0$ for the 350 galaxies SDSS sample matches the CF4 catalogue value. As the zero-point passes this test, this assures us that spurious large scale flow fields are not added to the CF4 dataset. 

\subsubsection{Identifications from DELVE DR2 alone}
In the previous sections we have used the positional information from identifications made by \citep{2024SCPMA..6719511Z} to cross
match with optically catalogued galaxies. We now attempt to match some of the non-SGA and non-SDSS FAST 2023 sources
with DELVE DR2 galaxies.
We tested acceptance radii of 0.18 arcmin, %(black points in Figure 8)
0.24 arcmin  %(black open circles)  
and 0.27 arcmin. %(red open circles). 
In this mass-radius relation there is a tendency for spurious objects to fall below the lower envelope of the real ones. None of the chosen radii seem to fail this condition, however.

\begin{figure}
\centering
\includegraphics[width=\linewidth,angle=-0]{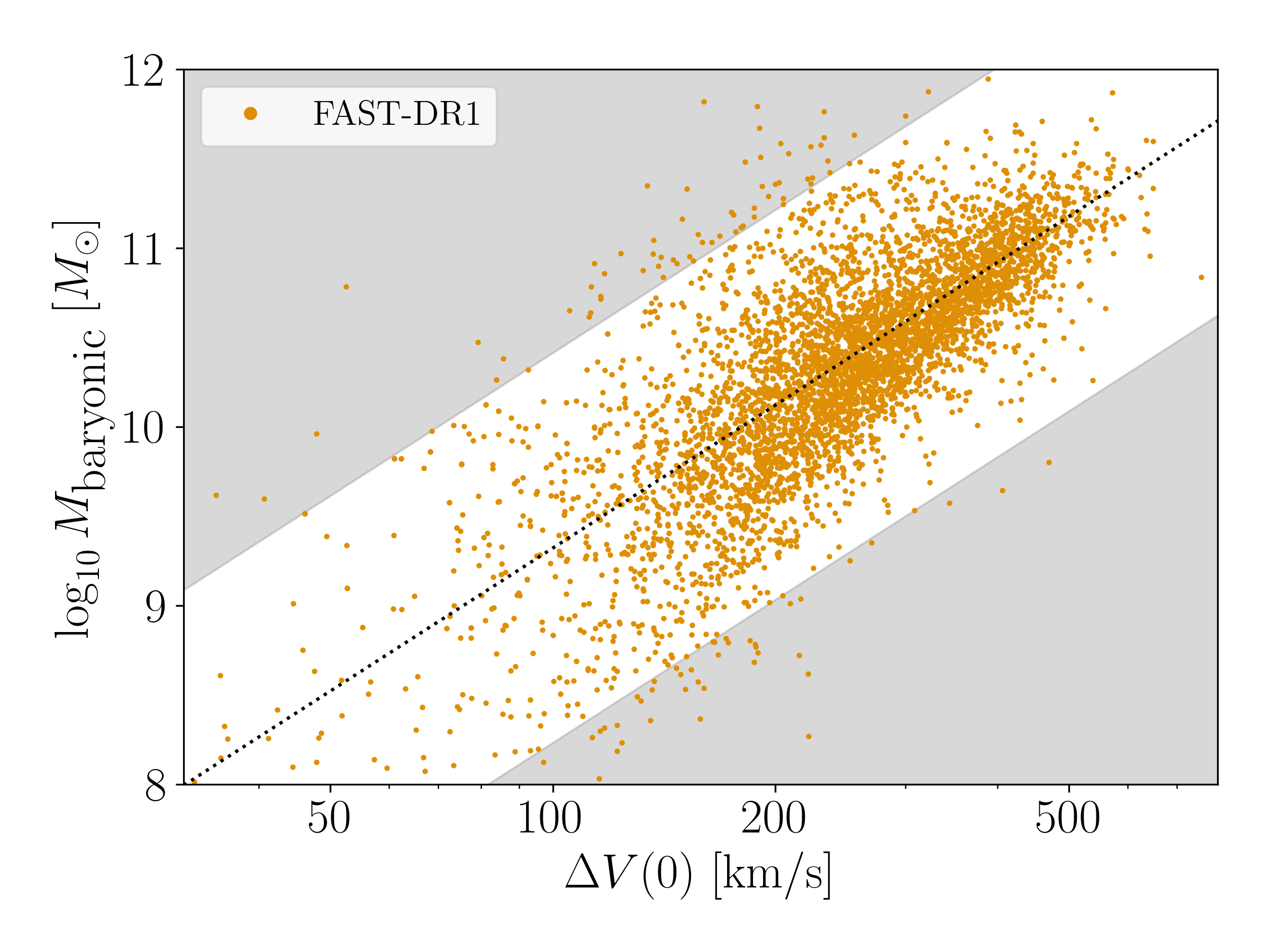}
\caption{The baryonic Tully-Fisher relation for the 4,796 FAST galaxies. The dotted line shows the fitted least-squares relation: $\log_{10}M_\text{baryonic}=10.27+2.65(\Delta V(0)-2.36)$. The galaxies located in the grey regions are omitted from the final sample, as they demonstrate deviations greater than 2.5$\sigma$ from the TFR fit.}
\label{fig:TFR}
\end{figure}

 The DESI Legacy Survey\footnote {\url{www.legacysurvey.org}} as part of DR9 has provided Tractor photometry of northern fields, and we have matched these with FAST sources in the same way as for the DELVE survey. We included every object with $z< 22$ and half light radius greater than 3$^"$. The mass radius relation looks the same as for the DELVE identifications, with the same independence on matching radius out to 2.7 arcmin. We chose the zero-point of the Tractor photometry to match that of the SGA photometry. With over 2000 galaxies this was achieved to an accuracy of 0.005 mag.

One interesting feature of the DESI Legacy Survey tractor photometry is the provision of ellipticity uncertainties. This catalogue includes the real and imaginary weak lensing ellipticity components, $\epsilon_1$ and $\epsilon_2$. For gravitational micro-lensing analysis where $\epsilon ~=~ \surd (\epsilon_1^2~+~\epsilon_2^2)$ and 
\begin{equation}\label{eqn:ratio}
    b/a = \frac{1-\epsilon}{1+\epsilon} \\,
\end{equation}
where $b/a$ is the axial ratio, and are related to the inclination of galaxies via:
\begin{equation}
    \cos^2 i = \frac{(b/a)^2 - q_0}{1-q_0} \\, 
\end{equation}
and $q_0=0.2$ is the standardly adopted intrinsic axis ratio. By differentiating equation (\ref{eqn:ratio}) for $\cos i$ we obtain the fractional uncertainty in $\sin i$: 
\begin{equation}
    \frac{\delta \sin i}{\sin i} =
\frac{(1-\epsilon)^2}{4~\ln(10)\epsilon (1+\epsilon)} (\epsilon_1 \delta \epsilon_1 + \epsilon_2 \delta \epsilon_2) \\.
\end{equation}
For galaxies with $0.2~<~b/a~<~0.7$, $z~<$ 22  and half light radius exceeding 3$^"$ we obtain the baryonic Tully Fisher relation shown in Figure \ref{fig:TFR} . All but a few galaxies have $\log~\Delta$V(0) uncertainties from $\delta \epsilon_{1,2}$ of 0.01 dex, and the modal uncertainty is in the third decimal place. %The DESI Legacy survey is the  source of these uncertainties.

\subsection{DESI-PV data release 1}
\cite{2024arXiv240813842S} has released the first sample of peculiar velocities from the DESI survey. The peculiar velocities of galaxies were computed using the Fundamental Plane relation for early-type galaxies and the Tully-Fisher relation for late-type galaxies. During the DESI survey's validation phase, stellar velocity dispersion data were gathered for 6,698 early-type galaxies. Of these 4,191 were converted to CF4 calibration, none of which were in common with CF4, while a few were in common with FAST; their distances and uncertainties (the latter are similar to the FAST data) were accepted after setting little $h$  to agree with CF4's (their distances assume H$_0$ = 100 \kmsmpc).

\section{Reconstruction of the CF4++ gravitational velocity field}\label{sec:reconstruction}

\begin{figure}	 
\centering
\includegraphics[width=.8\linewidth,angle=-0]{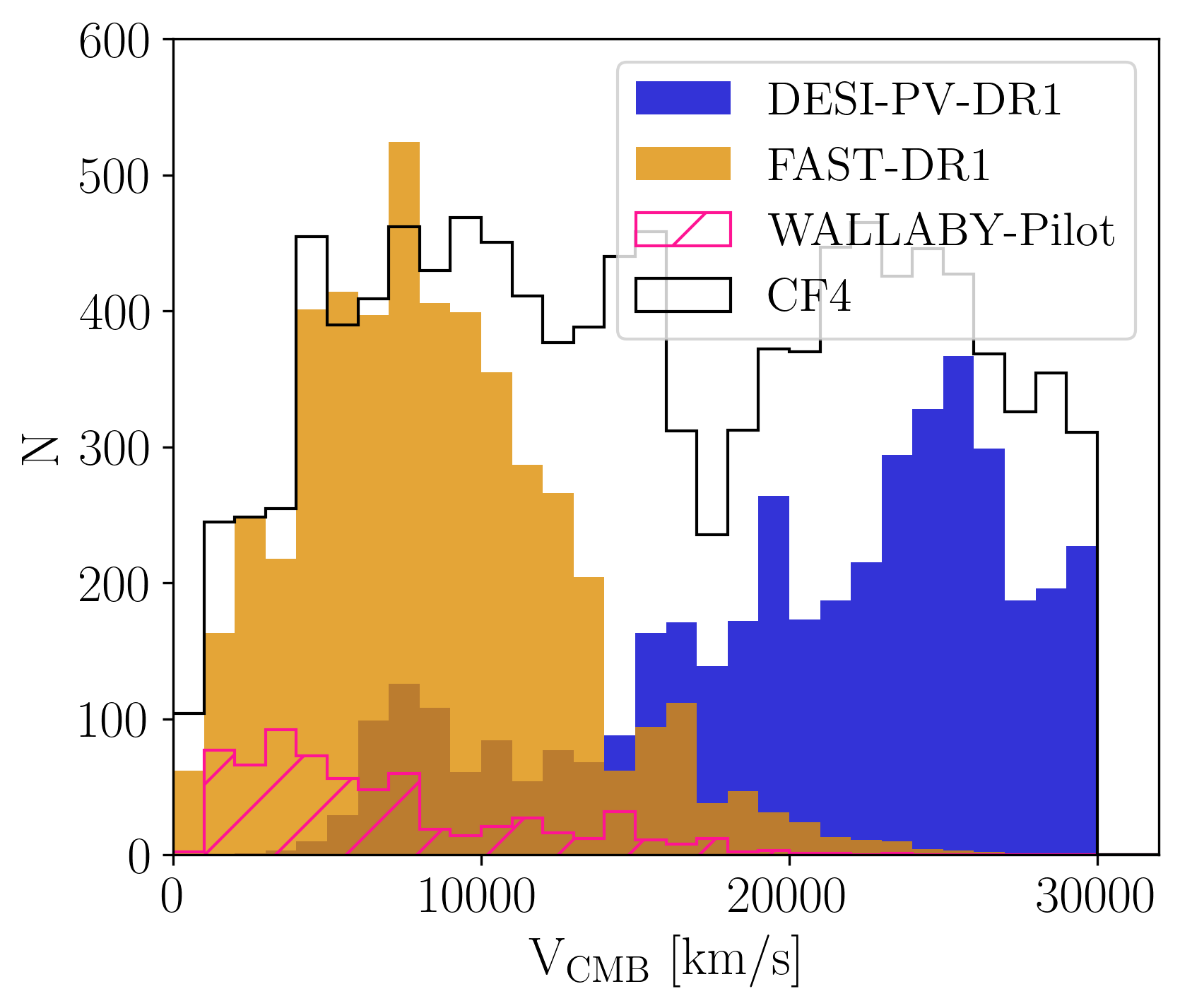}
\caption{Distribution in redshift of the four datasets used in this article in bins of 1,000 \kms, for easier comparison the number counts of the CF4 galaxies have been reduced by a factor of 5. FAST-DR1 and WALLABY pilot phase are probing the Universe up to 20,000 \kms, while DESI-PV-DR1 extends up to 30,000 \kms, as do the  SDSS galaxies that were incorporated in CF4.}
\label{fig:histogram_Vcmb}
\end{figure}

In this section, we update the local Universe velocity and full matter (dark+luminous) over-density fields and compare to the previous version using only CF4 datasets \citep{C23}.
The velocity fields are reconstructed in a self-consistent way, the inferred field at any location depend non-locally on all the other points in the dataset. The method used to calculate these fields is the same as those outlined in Section 4 of \cite{C23}, we refer the reader to there for details of how these calculations are performed. However, here we will clarify that the robustness of the reconstruction presented in this work is inherently linked to the observational errors, naturally preferring galaxies with smaller uncertainties. As a result, none of the datasets are inherently weighted more heavily, but the WALLABY and FAST galaxies do have on average lower uncertainties, and the method naturally prioritizes higher-quality data due to its Bayesian approach.

Figure \ref{fig:histogram_Vcmb} shows the distribution of the four datasets in redshift. One can see that FAST-DR1 and WALLABY pilot surveys are mapping the Universe out to velocities of 20,000 \kms, whereas DESI-PV-DR1  extends further out to 30,000 \kms, similar to the SDSS peculiar velocity data previously included in CF4.
For simplicity in the rest of the article,  the catalogue composed of CF4 with the new additions from WALLABY Pilot phase, FAST-DR1, and DESI-PV-DR1 will be referred to as CF4++.

\begin{figure}	
\centering
\includegraphics[width=\linewidth,angle=-0]{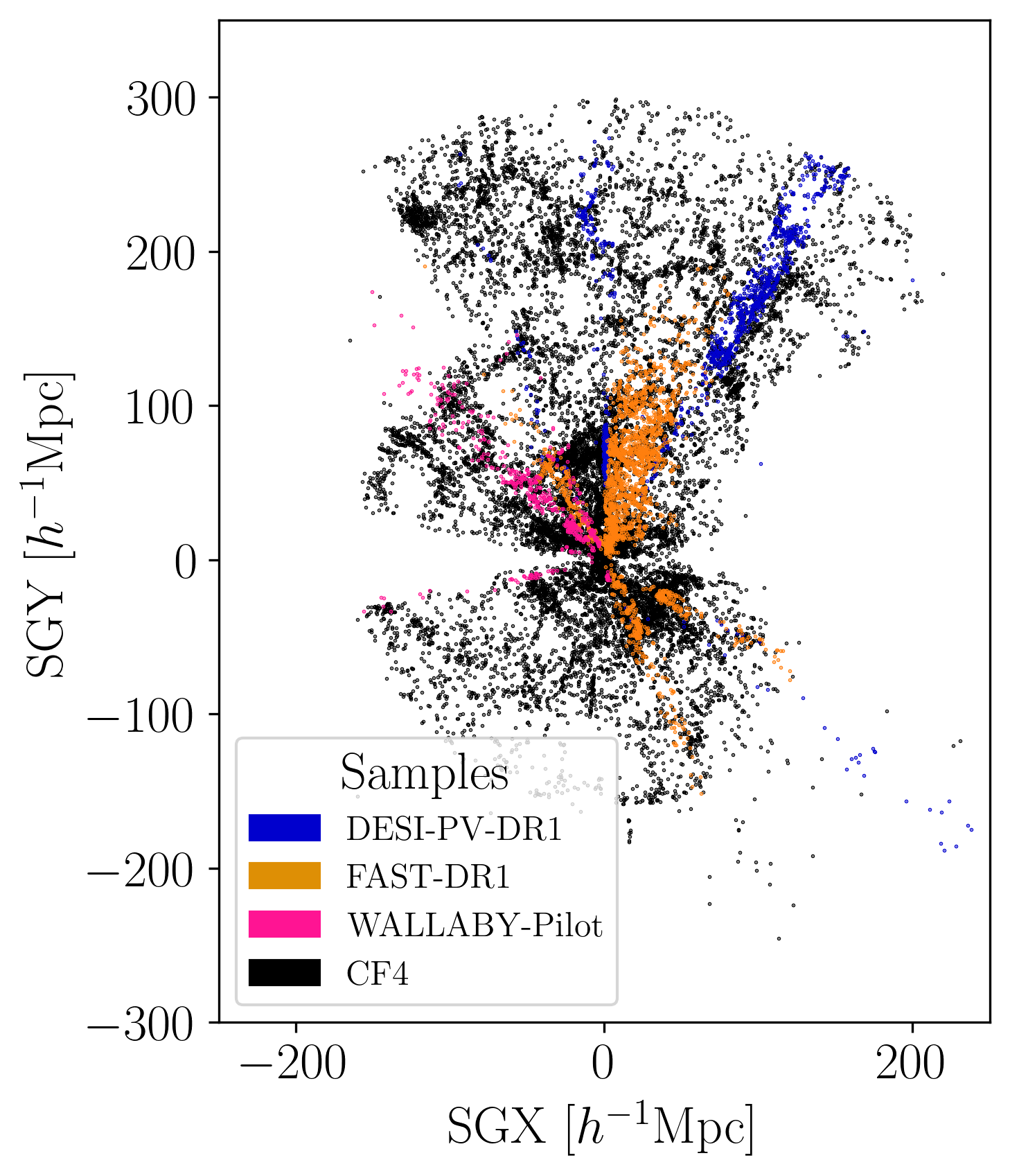}
\caption{The distribution of galaxies with measured peculiar velocities in the local Universe, within $z = 0.1$ and between -23.4 \h Mpc $<$ SGZ $<$ 23.4 \h Mpc. CF4 has been significantly enhanced by the addition of 9,642 galaxy distances. The WALLABY and FAST surveys are extending measurements to approximately 150 $h^{-1}$Mpc, while DESI-PV is expanding coverage to the SDSS limits in the northern terrestrial hemisphere.
}
\label{fig:SGX-SGY_data_1panel.png}
\end{figure}

\begin{figure}	
\centering
\includegraphics[width=\linewidth,angle=-0]{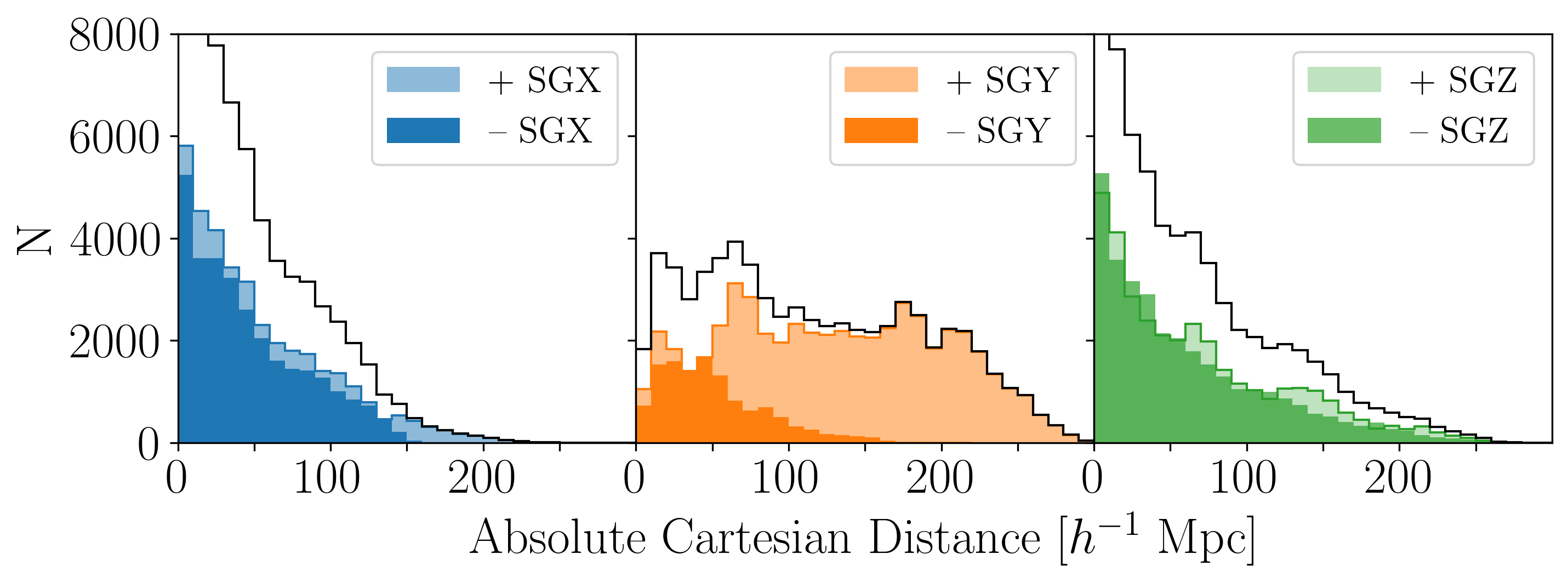}
\caption{The number of galaxies as a function of radial bins of size 10 \hmpc. The black histograms show the overall total of each of the Cartesian Supergalactic coordinates, while the light and dark histograms show the positive and negative components.  
\textcolor{blue}
}
\label{fig:cartesian_hist}
\end{figure}

Figure \ref{fig:SGX-SGY_data_1panel.png} displays the galaxies in CF4++, within SGZ$\leq\pm 23.4$ \hmpc, in SGX-SGY coordinates. The original Cosmic-Flows 4 catalogue data is shown by the black points.  The WALLABY Pilot phase, FAST-DR1 and DESI-PV-DR1 galaxies that were integrated into CF4++ are shown as coloured circles. 
The histograms of Figure \ref{fig:cartesian_hist} illustrate the absolute distribution of the CF4++ Cartesian Supergalactic coordinates. As depicted, CF4++ exhibits a relatively symmetrical distribution in both the SGX and SGZ directions. However, it shows a significant preference for the North (+SGY), with only 10\% of galaxies extending into the -SGY coordinates.
One can see that the current compilation of galaxy peculiar velocities is highly inhomogeneous, both in sky coverage and data quality. The northern extension of CF4, based on the SDSS-PV sample, is particularly affected by distance-dependent errors, necessitating caution when interpreting flows in that region.

\begin{figure}	 
\centering
\includegraphics[width=\linewidth,angle=-0]{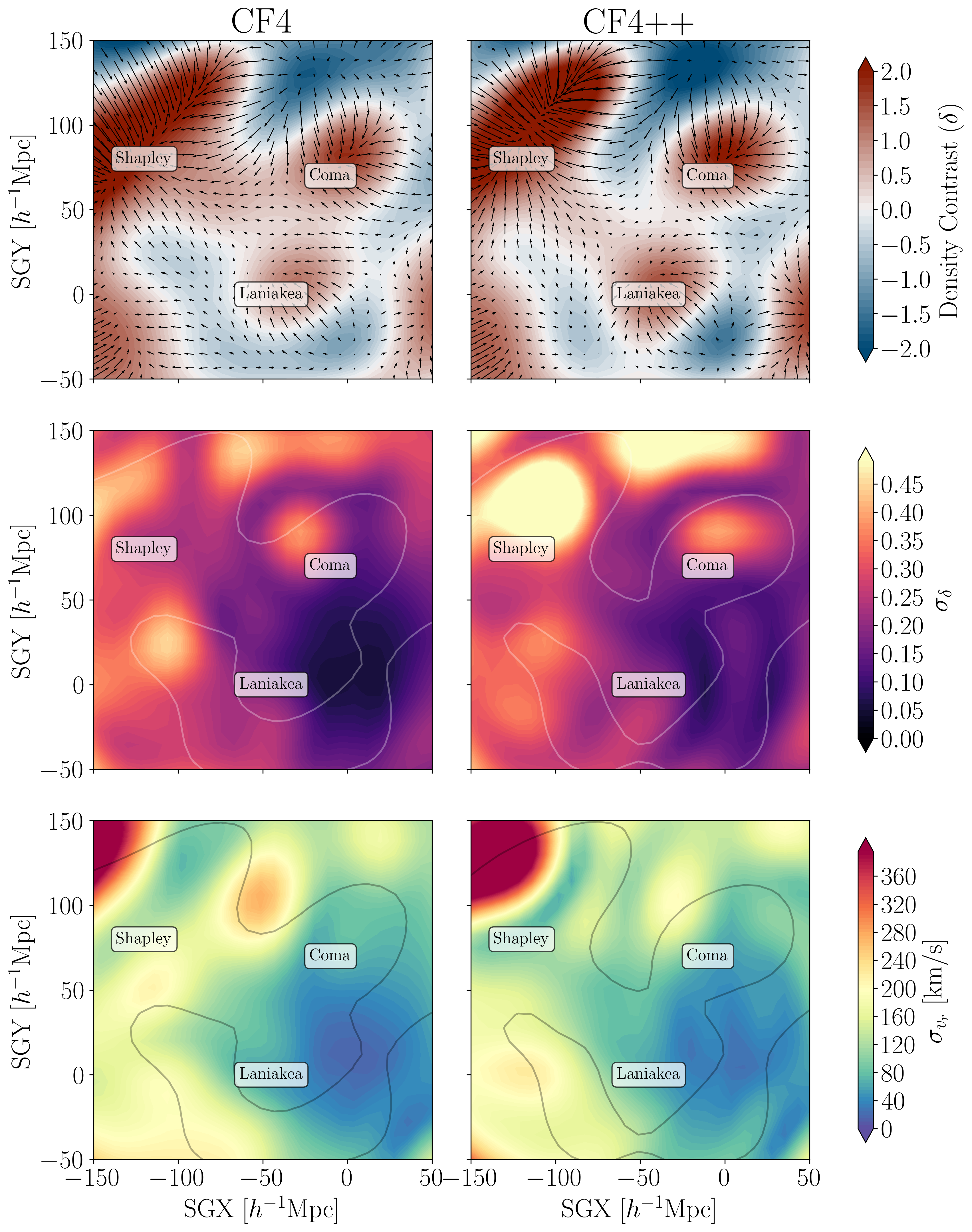 }
\caption{The CF4 (left) and CF4++ (right) local cosmography centred on SGZ = -3.9  \h  Mpc, averaged over a total width of 24.3 \h Mpc. The first row shows the mean of the HMC realizations density contrast ($\delta$) and is over-plotted by the mean radial velocity field. 
The second and third row respectively show the standard deviation of the density contrast fields, and the radial velocity fields. The grey contours over-plotted on these slices are the equivalent contours lines of where $\delta=0$. }
\label{fig:std}
\end{figure}

Figure \ref{fig:std}
focuses on the nearby Universe, in particular on Coma cluster seen at SGX-SGY=(0;60) $h^{-1}$Mpc and the Great Attractor region at SGX-SGY=(-45;0) $h^{-1}$Mpc. The updated cosmography clearly delineates Coma as a stand-alone large scale structure not connected to the Shapley region. The Great Attractor (Laniakea) is reinforced as a major contributor to the local density field. 
This figure additionally illustrates the uncertainty in the density and velocity fields for this region. Here we show the standard deviation for both, calculated using all our Hamiltonian Monte Carlo (HMC) solutions. One can see that while there is some variation in the upper structure of Shapley and the nearby void region, the favoured over-density regions of Coma and Laniakea are fairly stable. 

To ensure reliability, we provide users with an estimate of uncertainty by calculating how much the predicted values vary over the 10,000 different steps in our sampling process. Providing a better understanding for users on the stability of the reconstruction at any specific location.

\begin{figure}	
\centering
\includegraphics[width=\linewidth,angle=-0]{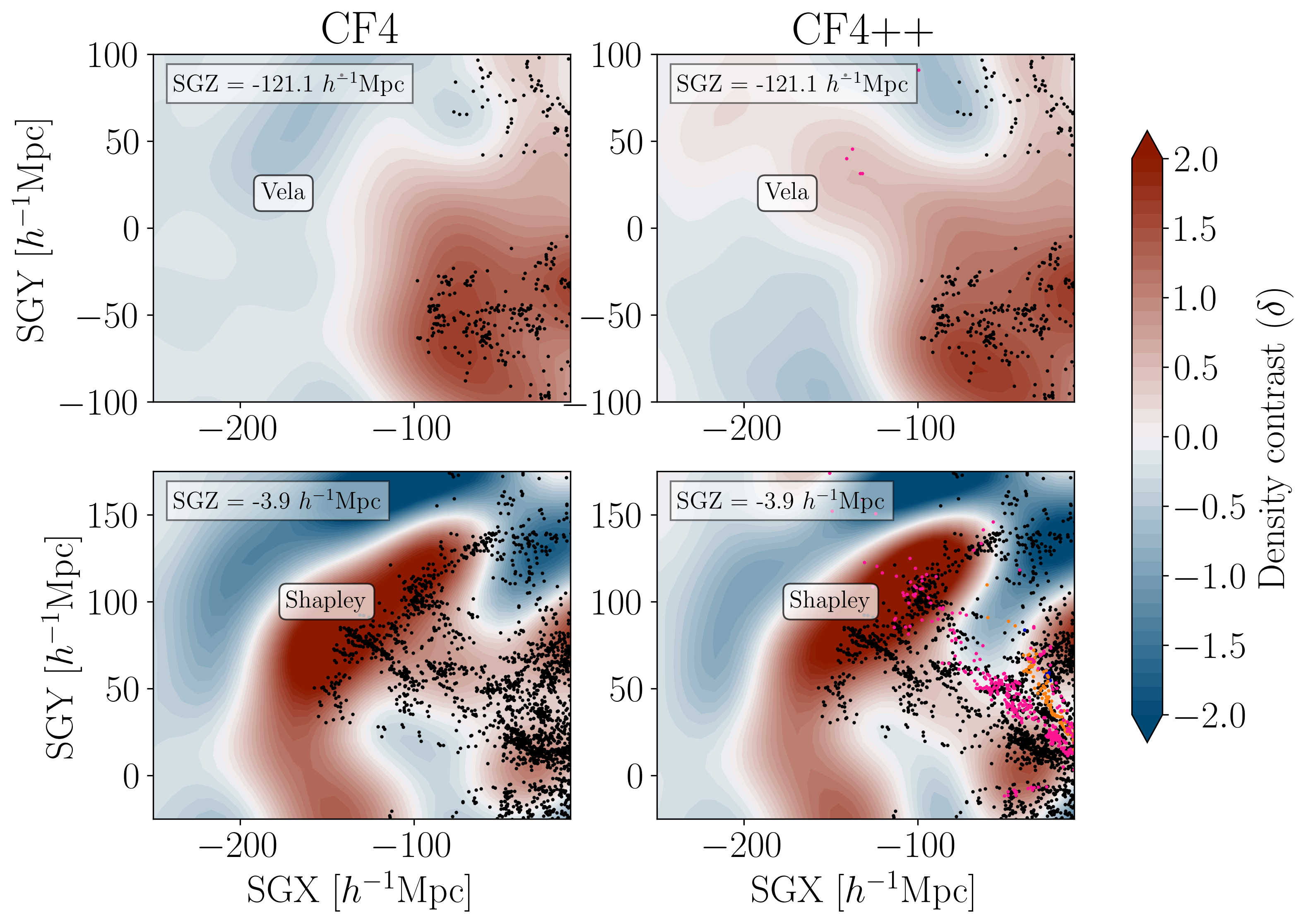}\\
\caption{The reconstructed matter density field, focusing on the Vela (top) and Shapley (bottom) supercluster regions, centred on SGZ = -121.1 \h Mpc and -3.9 \h Mpc respectively, each averaged over a total width of 23.4 \h Mpc. The left panel displays results derived from the CF4 dataset, while the right panel uses the CF4++ dataset. Black dots represent CF4 galaxies, and the coloured dots correspond to newly added data points located within the averaged region.}
\label{fig:SGX-SGY-VELA_Density_slice_2panels}
\end{figure}

Figure \ref{fig:SGX-SGY-VELA_Density_slice_2panels} focuses on Vela (top) and Shapley (bottom)  supercluster regions by presenting the reconstructed matter density contrast based on the CF4 dataset (left) and the CF4++ dataset (right). The CF4 galaxies are marked by black dots, while newly incorporated data points are shown as coloured dots. The addition of only very few measurements near the galactic plane reveals the emergence of the Vela supercluster. The additional data also modifies the density contours of Shapley, both in its foreground and background. Shapley is clearly separated from the foreground cluster Coma, while Laniakea remains connected to and moving towards Shapley. Future surveys and continued observations in this part of the sky will refine our understanding of these massive structures and their significance for near-field cosmology.

\section{Bulk flow and homogeneity scale}\label{sec:results}

In order to analyse the bulk flow, we first have checked that the dataset was not biased towards a specific environment of the V-web.

The V-web is a method for identifying the cosmic web's structure—voids, walls, filaments, and nodes— using the velocity field of galaxies or matter. It was introduced as an alternative to density-based methods in the context of Local Universe peculiar velocity studies in \cite{2014Natur.513...71T, 2015MNRAS.448.1767C}, as well as more recently in the WALLABY survey \cite{2021MNRAS.507.2300F, 2024MNRAS.533..925M}, and in the study of cosmic voids by the CAVITY consortium \cite{2023A&A...673A..38C, 2024A&A...687A..98C}.
 The approach focuses on computing the eigenvalues of the velocity shear tensor, derived from the peculiar velocity field. The tensor captures how velocities change across a region, reflecting the large-scale gravitational dynamics.

By classifying regions based on the eigenvalues, the V-web assigns them to one of the four cosmic web structures. This method emphasizes the dynamical state of the cosmic web rather than static density contrasts, offering a more physically grounded classification. It is particularly advantageous for studying the dynamics of large-scale structure formation and the interplay between matter and gravity over cosmic time.

\begin{figure}
\centering
\includegraphics[width=\linewidth,angle=-0]{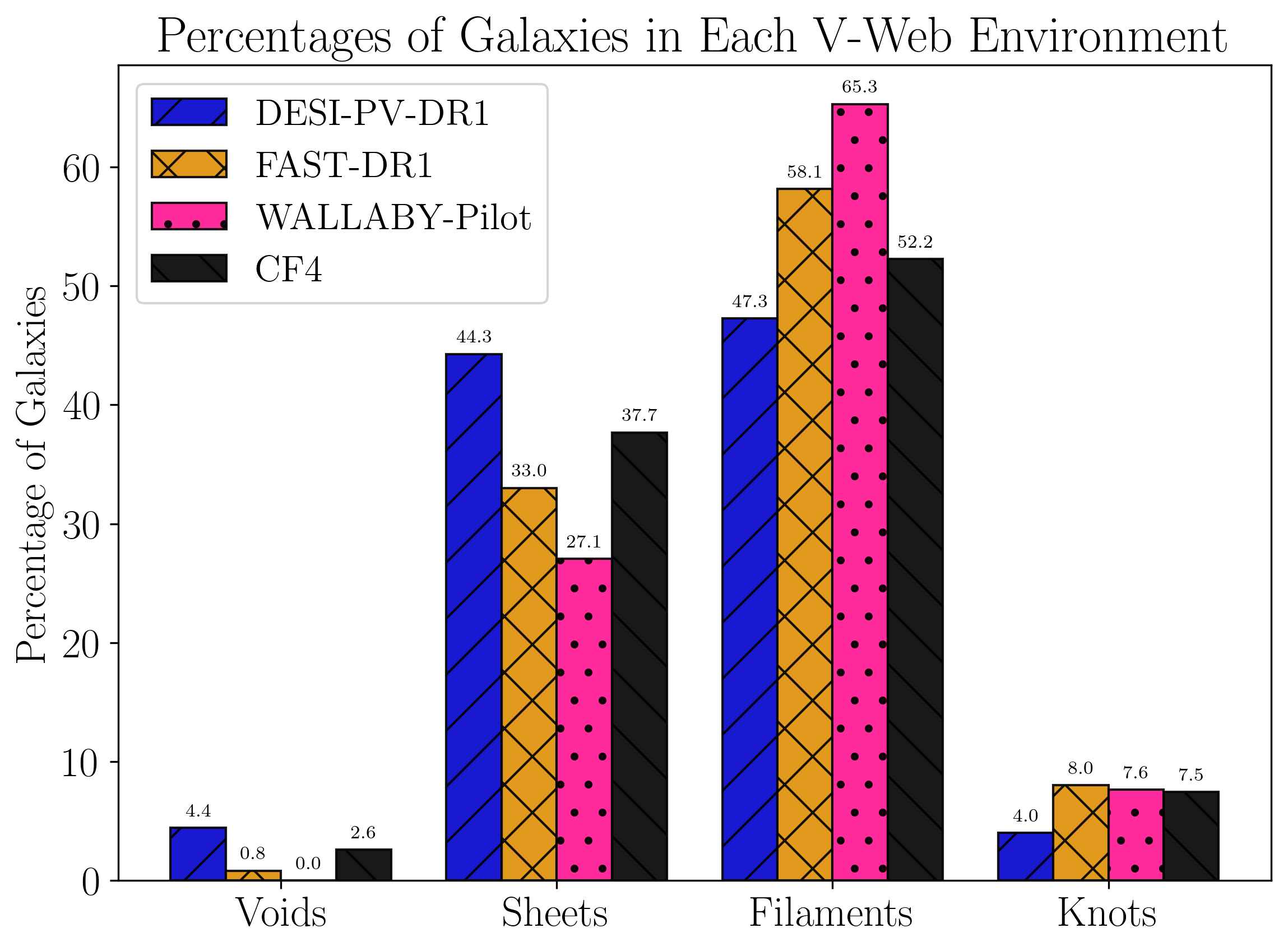}
\caption{Distribution of CF4++ galaxies in cosmic-web environments. WALLABY was targeted towards clusters of galaxies, while FAST-DR1 and DESI-PV-DR1 are representatives of all cosmic web environments.}
\label{fig:histogram_vweb}
\end{figure}

Figure \ref{fig:histogram_vweb} shows the distribution of the galaxies in the four different V-web settings. Since the Pilot phase of WALLABY focused on galaxy clusters, it does not include galaxies located in cosmic void environments. In contrast, the FAST-DR1 and DESI-PV-DR1 datasets include galaxies spanning all classical cosmic web environments, making them representative of the full range of large-scale structures in the Universe.

\subsection{Determining the bulk velocity}
As peculiar velocity surveys extend to larger scales, it is expected that the overall bulk flow within the observed volume will tend toward zero when compared to the reference frame established by the cosmic microwave background. The extent to which this bulk flow diminishes in relation to survey volume serves as an indicator of large-scale modes of the matter power spectrum and density field.
Given that the velocity fields are evaluated on a 3D Cartesian grid, the bulk velocity components can be easily determined as the volume-weighted average velocity within a top-hat sphere of radius R as follows:
\begin{equation}
    \boldsymbol{V}_{\textrm{bulk}}(R) = \frac{3}{4 \pi R^3} \int_{x<R} \boldsymbol{v} (\boldsymbol{x})\textrm{d}^3x \\.
\end{equation}

\begin{figure}
\centering
\includegraphics[width=\linewidth,angle=-0]{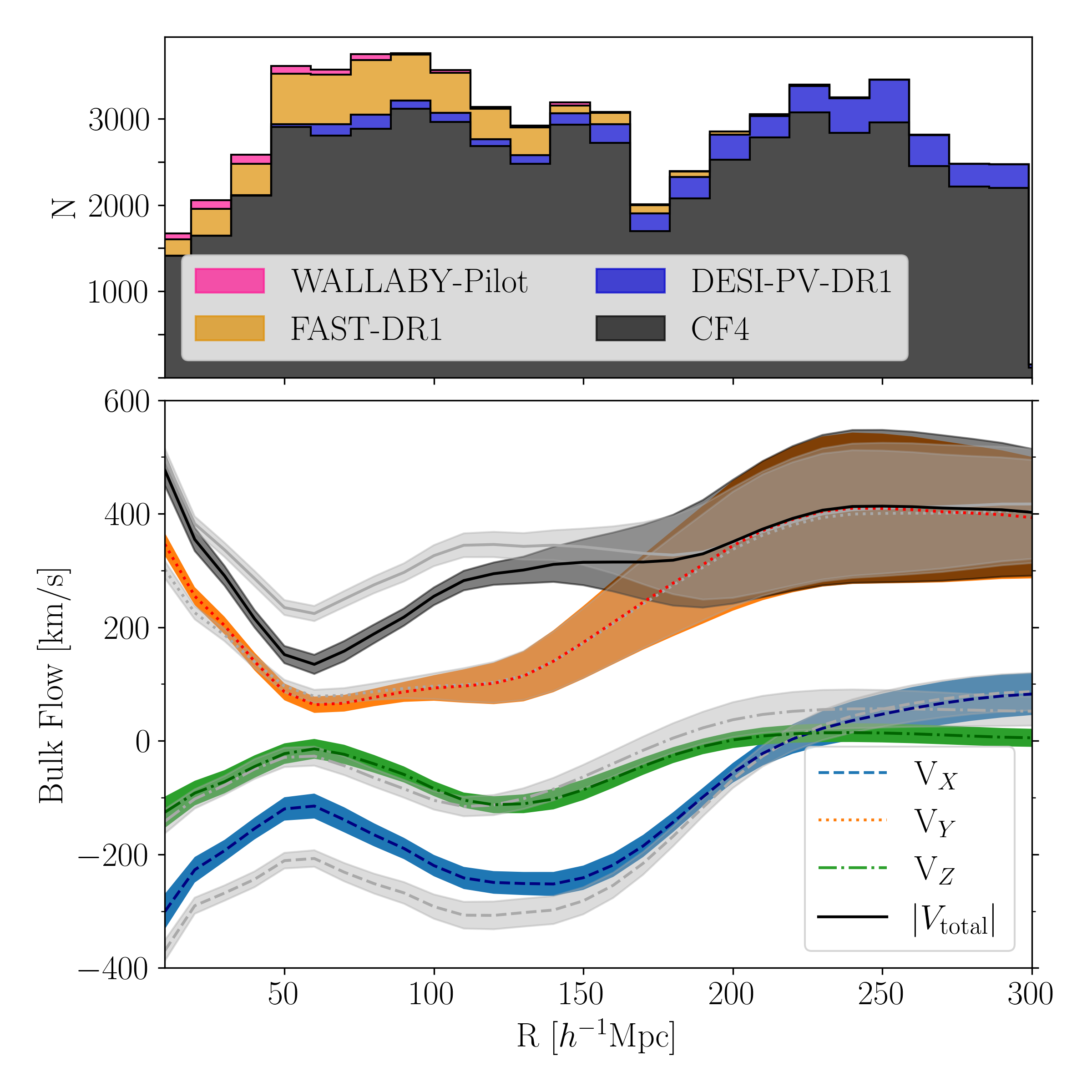}
\caption{(Top) Stacked histogram showing the radial distribution of the galaxies composing the CF4++ dataset  and (bottom) the mean bulk flow components along the Supergalactic X (dashed) , Y (dotted), and Z (dash-dotted) axis and the total bulk flow (solid), as a function of distance from the observer's location. The coloured bands show the CF4++ results, while the light grey bands show the CF4 measurements. The mean and standard deviation (transparent bands) are calculated using the nearly 10,000 HMC realizations. }
\label{fig:bulkflow}
\end{figure}

Figure \ref{fig:bulkflow} presents the amplitude of the total bulk flow (solid line) as a function of distance from the observer's position, along with the three Cartesian components (dashed lines) along the Supergalactic X, Y, and Z axes. The top panel shows a stacked histogram of the CF4++ data sets, while the bottom panel show the mean and standard deviation of our CF4++ and CF4 HMC realizations. This illustrates how the datasets introduced to CF4 influence each component and the overall bulk flow measurements. The SGY bulk flow component exhibits increasingly large fluctuations and does not converge to zero beyond a distance of 150 \hmpc. In contrast, the SGX and SGZ components begin to approach zero at this scale. It is important to recall that convergence to zero in the reconstruction indicates the absence of additional data and suggests that the reconstructed velocity field aligns with the zero-velocity solution predicted by $\Lambda$CDM, corresponding to the Universe's mean density.
However, Figures \ref{fig:SGX-SGY_data_1panel.png} and \ref{fig:cartesian_hist} illustrate that there are minimal additional data available beyond 150 \hmpc\ in the SGX and SGZ directions, whereas the SGY direction still contains a tail of data derived from SDSS Northern peculiar velocity measurements. These SDSS data are highly biased and uncertain, resulting in larger scatter in the HMC solutions, as evidenced by the broader width of the orange curve in Figure \ref{fig:bulkflow}. 

%as expected at larger distances, with increasing homogeneity of the Universe, the bulk flow tends towards to zero. 

%As illustrated in Figure \ref{fig:bulkflow},

Our 3D HMC realizations are significantly affected by abrupt declines in the number of galaxies observed, as these reductions can lead to incomplete or biased sampling of the cosmic structure. We find that radii associated with sharp decreases in galaxy counts correspond to the radii with increased variability in the velocity fields, and hence our recovered bulk flows.  This is especially true for the SGY-component of the bulk flow as the data is heavily skewed in favour of the +SGY component, as seen in Figure \ref{fig:cartesian_hist}, whereas both the SGX and SGZ components are significantly more symmetrical in distribution and trail off at $\sim$ 200 and 250 \hmpc. This imbalance in the current dataset is hence reflected in both our total and Cartesian bulk flow measurements and the amount of standard deviation therein.  In particular despite the large number of galaxies located around SGX, SGY $\approx$ -110, 220 (part of the Sloan Great Wall) which can be seen in Figure \ref{fig:SGX-SGY_data_1panel.png}, the radial velocity field in this region indicates that there is potentially a large structure beyond the current survey limits. This region is partially responsible for driving  the large bulk flow measurements we find at distances greater than $\sim200$\hmpc. This discrepancy in data means that we do not know if there is comparable structure in the South that would mitigate the current observed excess flow.  %In figure~\ref{fig:bulkflow_w_values} the benchmark bulk flow is predicted by the LCDM Quijote simulations \citep{Quijote20}.

\subsection{Comparison of recent measurements with $\Lambda$CDM predictions}\label{sec:comparison}

As the amplitude of the bulk flow, $|\boldsymbol{\mathrm{B}}|$, is sensitive to the matter power spectrum's large-scale modes, for a given cosmological model the measured bulk flow can thus be compared with the predicted value. The expected mean of the bulk flow is zero, as statistically the Universe is homogeneous and isotropic. However, the root-mean-square variance of the bulk flow amplitude, provides insight into our local Universe as it is dependent on the matter power spectrum $P(k)$, the scale $R$, as well as the window function $W(R)$ in which it is measured. 

Over the past 15 years, various studies have investigated the bulk flow. We present a summary of these values and their quoted distances in Table \ref{table:values}. %and graphically in Figure \ref{fig:bulkflow_w_values}.
While some studies report values exceeding those predicted by  $\Lambda$CDM \citep[notably,][recently]{W23,WH23}, others find no significant differences from expectations. 
Assuming linear perturbation theory, the variance of the bulk flow at a given scale, is given by:
\begin{equation}
    \sigma^2_\textrm{BF}(R) = \left< \boldsymbol{\mathrm{B}}^2 \right> = \frac{H_0^2 f^2}{2 \pi^2} \int_0^\infty P(k) \tilde{W}^2(k;R)  \textrm{d}k \\,
\end{equation}
where $\tilde{W}(k)$ is the Fourier transform of the window function \citep{gorski_pattern_1988} and $f 
\approx \Omega_m^{0.55}$ is the present-day growth rate of cosmic structure assuming a $\Lambda$CDM cosmology. In this work, we use the publicly available python CAMB package to calculate the non-linear matter power spectrum \citep{Lewis_camb_2000} using the \cite{planck_collaboration_planck_2020-values} cosmology ( $\Omega_\Lambda = 0.685, \Omega_m= 0.315, h = 0.674, \sigma_8 = 0.811$, and $n_s = 0.965$),   and assuming a spherical top-hat window function, such that $\tilde{W}(k;R) = 3(\sin(kR)-kR\cos(kR))/(kR)^3$.

Assuming the density field is Gaussian, the peculiar velocity field is then given by the Maxwellian distribution and the PDF can be estimated using \citep{Bahcall_1994,Li_2012}:
\begin{equation}\label{eqn:prob}
    p(|\boldsymbol{\mathrm{B}}|)dB= \sqrt{\frac{2}{\pi}}\left(\frac{3}{\sigma_{\textrm{BF}}^2}\right)^{3/2}|\boldsymbol{\mathrm{B}}|^2\exp \left(-\frac{3|\boldsymbol{\mathrm{B}}|}{2\sigma_{\textrm{BF}}^2} \right)dB   \\,
\end{equation}
where the most probable $\boldsymbol{\mathrm{B}}$ is $B=\sqrt{2/3}\sigma_{\textrm{BF}}$, and the 1$\sigma$ (68\%) and 2$\sigma$ (95\%) confidence levels used in this work are calculated by integrating equation (\ref{eqn:prob}), but are approximately  $B^{+0.419\sigma_{\textrm{BF}}}_{-0.356\sigma_{\textrm{BF}}}$ and $B^{+0.891\sigma_{\textrm{BF}}}_{-0.619\sigma_{\textrm{BF}}}$ \citep{S16}. %respectively shown as the light and dark colored bands in this figure.

\begin{figure}
\centering
\includegraphics[width=\linewidth,angle=-0]{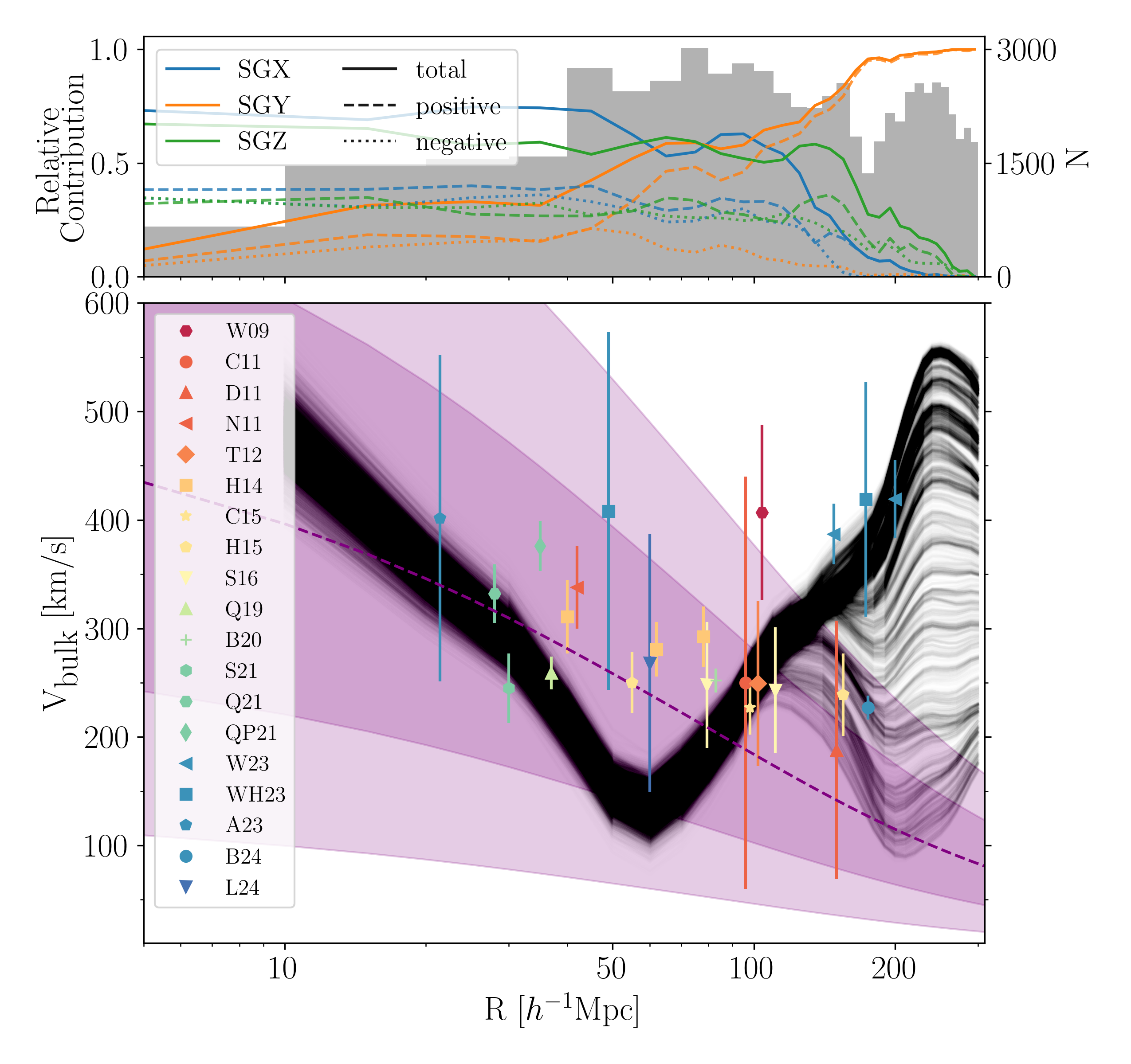}
\caption{(Top) The solid (dashed, dotted) lines show the absolute total (positive, negative) relative contribution (the square root of the fractional contribution) of each Supergalactic Cartesian coordinate to the overall CF++ radial distribution; while the histogram shows the total number of galaxies in bins of 10 \hmpc. (Bottom) The total bulk flow of each HMC CF4++ realization (black), as a function of distance. Over-plotted are the Table \ref{table:values} values assuming an effective radius corresponding to a spherical top-hat window function, and the dashed curved line is the corresponding linear theory  $\Lambda$CDM prediction assuming the Planck cosmology specified in Section \ref{sec:comparison}. 
The light and dark regions represent the $1\sigma$ and $2\sigma$ confidence levels, respectively.  Small offsets have been introduced in the cases of overlapping data points.}
\label{fig:bulkflow_w_values}
\end{figure}

To accurately compare measurements from specific datasets with the theoretical predictions of  $\Lambda$CDM,  requires one to account for any potentially complex survey window function. However, to facilitate comparisons across different datasets, it is essential to standardize this window function. In Figure \ref{fig:bulkflow_w_values}, we compare our bulk flow amplitude of each CF4++ HMC realization as a function of distance from the observer, against previous measurements, and the $\Lambda$CDM prediction assuming a spherical top-hat window. The top panel illustrates not only the overall CF4++ distribution, but also the relative contribution each Supergalactic Cartesian coordinate has on the overall distance. This additionally illustrates that the increase in the scatter between realizations correlates with where there is an overall decrease in data, and where the galaxies overall positions are dominated by the +SGY direction.
The values quoted in Table \ref{table:values} that were calculated using an alternative window function, are adjusted to a new effective radius ($R_e$), following the methodology in \cite{S16}.    
Measurements, which were calculated based on a Gaussian filter,  are plotted at $R_e=2R$ to align them more closely with the predictions of the spherical top-hat window. Similarly, the values presented in \cite{S16} use a Gaussian filter and are based on the 6dFGSv survey which is akin to a hemispherical top-hat, so following their arguments we plot their quoted values at $R_e = 2(R^3/2)^{1/3}$. If the methodology used to calculate the effective radius of the bulk flow measurement was unstated in the original paper, we 
make no corrections and measurements are placed at their reported bulk flow depths.

In regards to these previous studies, we can distinguish no significant trend for the recovered bulk flow, neither in terms of recency of the study nor the data used in each measurement. We find that the majority of our bulk flow measurements are consistent with most previous works quoted values at comparable radii, with the exception of the region between $\sim$ 50-100 \hmpc\, here we recover a bulk flow that is significantly lower than previous measurements.  We attribute this discrepancy to the inclusion of the FAST dataset, which reduces the SGX component of the bulk flow by $\sim$ 100 \kms at these depths, as illustrated in Figure \ref{fig:bulkflow}. We also recover a lower measurement of the bulk flow of $315 \pm 40$ \kms\ within 150 \hmpc, when compared to the previous CF4 dataset which yields $342 \pm 32$ \kms.  

Both our and previous bulk flow measurements exhibit no significant tension with  $\Lambda$CDM predictions within 150 \hmpc.  Within 100 \hmpc\ we find no discernable tension between the recovered CF4++ and the $\Lambda$CDM predicted bulk flows.  Between 100-150 \hmpc\ we find a moderate tension averaging on the order of 1.7$\sigma$ with $\Lambda$CDM.  Extending beyond 150 \hmpc, we see not only increasing fluctuations in the HMC realizations, but also an increasing tension between  $\Lambda$CDM predictions and CF4++ measurements, achieving a maximal tension of 3.0$\sigma$ at 300 \hmpc.

\begin{table*}
\begin{tabular}{llllll}
  \hline
Work : Selection Function  & Dataset(s) & \multicolumn{3}{l}{Measurement within Distance Range [\kms]:}  \\% & \\
& &  0-50 $h^{-1}$Mpc & 51-100 $h^{-1}$Mpc & 101-150 $h^{-1}$Mpc & 151-200 $h^{-1}$Mpc \\  \hline   

   \smallskip
   
    $\Lambda$CDM prediction ($\sigma_{BF}$) & & $ 476^{+258}_{-220} - 258^{+133}_{-112}$  & $ 256^{+132}_{-112}- 184^{+94}_{-80}$  & $ 183^{+94}_{-80} - 142^{+73}_{-62}$  & $ 141^{+72}_{-62} - 115^{+43}_{-36}$ \\% & \\
    \hline
   
W09 : Gaussian  & SFI++, SN, etc. & 407 $\pm$ 81 (50) &   & & \\ %&  1.79\\
%& \\according to C11 this is never stated in the text)
C11 : Top-Hat   & Union 2 SN &   & 250 $\pm$ 190 (100) &  & \\% & 0.31 \\
D11 : Top-Hat  & Union 2 SN & &  & 188 $\pm$ 119 (150)& \\%& 0.33\\
N11 : Top-Hat & SFI++   & 333 $\pm$ 38 (40)& & & \\%& 0.38\\
T12 : Gaussian & A1SN & 249 $\pm$ 76 (50)& & & \\%& 0.54\\
M13 : Top-Hat &  SFI++, A1SN, etc.  & 310 (50)& & & \\%& - \\
%H14\citep{H14} & 311 $\pm$ 334 (20), 281 $\pm$ 25 (30), 292 $\pm$ 28 (40)& & & \\
H14 : Gaussian & 2MTF &  292 $\pm$ 28 (40)& & & \\%&  0.2/0.33/0.76\\
C15 : Gaussian & 2M++ & 227 $\pm$ 25 (50) & & & \\%& 0.44\\ % I asked Mike and this was the correct value
H15 : Top-Hat & CF2 & 250 $\pm$ 21 (50) & & 239 $\pm$ 38 (150) & \\%&  0.06/1.19 \\
S16 : Hemisphere & 6dFGSv  & 248 $\pm$ 58 (50) &  243 $\pm$ 58 (70) & &  \\%& 0.3/0.66 \\
Q19 : Top-Hat & 2MTF, CF3 & 259 $\pm$ 15 (37) & & & \\%&  0.20 \\
B20 : Gaussian & A2SN,  SFI++ &252 $\pm$ 11 (40) & & & \\%& 0.41\\
S21 : Top-Hat & DSS &245 $\pm$ 32 (30) & & & \\%& 0.40 \\
Q21 : Top-Hat & 2MTF & 332 $\pm$ 27  (30) & & &\\%& 0.14 \\
QP21 : Top-Hat & CF4TF  & 376 $\pm$ 23 (35) & & & \\%& 0.53 \\
A23 : Top-Hat & ALFAFA & 401 $\pm$151 (21) & &  & \\%& 0.26\\
W23 : - & CF4  & &  & 395 $\pm$ 29 (150) & 427 $\pm$ 37 (200) \\%& 3.1/4.4 \\
WH23 : Top-Hat & CF4 & 408 $\pm$ 165 (49) & &  & 428 $\pm$ 108 (173) \\% & 0.69/2.3\\
B24 : - & CF4 &   & &   & 227 $\pm$ 11 (175)   \\
L24: -  &  SN-Pantheon+&   & 268 $\pm$ 119 (61) &  & \\%&  0.17\\
\textbf{This Work: Top-Hat} & \textbf{CF4++} &  \textbf{152 $\pm$ 15 (50)}  & \textbf{254 $\pm$ 16 (100)} & \textbf{315 $\pm$ 40 (150)} & \textbf{351 $\pm$ 109 (200)}  \\
\hline
Average Bulk Flow and & & $324 \pm 28 (37)$ & $275 \pm 87 (84)$ & $264 \pm 70 (140)$ & $355 \pm 66 (183)$ \\ 
Deviation from $\Lambda$CDM & & 0.21 $\sigma$  & 0.53 $\sigma$ &  1.12$\sigma$  & 2.54$\sigma$   \\
Average CF4++ and & & $298 \pm 20 (30)$ & $191 \pm 16 (80)$ & $301 \pm 28 (130)$ & $326 \pm 83 (180)$ \\ 
Deviation from $\Lambda$CDM & & 0.07 $\sigma$  & 0.16$\sigma$ &  1.70$\sigma$  & 1.93$\sigma$   \\
%average $\sigma$ dev & & 0.31 (0-50) & 0.71 (51-100) &  1.28 (101-150) & 2.22 (150-210) &  \\
%quad $\sigma$ dev & & 0.37 & 0.87 &  1.55 & 2.61 &  \\
\end{tabular}
\caption{Bulk Flow measurements in the literature using supernovae or galaxy peculiar velocity datasets (W09: \cite{W09}, C11: \cite{C11},D11: \cite{D11},N11: \cite{N11}, T12: \cite{T12},M13: \cite{M13},H14: \cite{H14},C15: \cite{C15},H15: \cite{H15},S16: \cite{S16},Q19: \cite{Q19},B20: \cite{B20},S21: \cite{S21},Q21: \cite{Q18},QP21: \cite{Q21},A23: \cite{A23},W23: \cite{W23},WH23: \cite{WH23},B24: \cite{B24},L24: \cite{L24}). The radii, listed in brackets, are the stated values of their respective articles in units of $h^{-1}$Mpc. Those with selection functions which are \emph{non} Top-Hat, when plotted in Figure \ref{fig:bulkflow_w_values}, are modified using the methodology described in Section~\ref{sec:results}. The $\Lambda$CDM and $1
\sigma$ confidence values quoted at the top of the table are the measurement ranges expected in their stated distance intervals. The bottom four rows of this table, respectively, calculate within each 50 \hmpc\ interval: 
(1) the average bulk flow measurement across the literature (BF$_{lit}$)  (assuming Top-Hat radii); (3) the average CF4++ measurement (BF$_{CF4++}$); respectively, the tension of (2) the BF$_{lit}$ and (4) the BF$_{CF4++}$ measurement with the $\Lambda$CDM prediction at the listed effective radius.}
%(1) the average bulk flow measurement within each bin across the literature (assuming Top-Hat radii); (3) the average of the five CF4++ measurements performed within each bin; (2) and (4) the  effective tension of these respective average bulk flow measurements calculated from $\Lambda$CDM at the effective radius.}
\label{table:values}
\end{table*}

\section{Conclusions}\label{sec:conclusion}

This study provides an updated cosmography of the local Universe within $z = 0.1$, incorporating new galaxy peculiar velocity datasets from preliminary data-releases of the WALLABY, FAST, and DESI surveys. This enhanced dataset, referred to as CF4++, includes 9,642 additional galaxy distances, allowing for a more detailed analysis of the local Universe's bulk flow dynamics and structure.

Key conclusions from the study include:
\begin{itemize}
    \item The bulk flow, which represents the coherent motion of galaxies due to gravitational interactions, was analysed using the CF4++ dataset.
    \item The updated cosmography reveals more spherical shapes and clearly defined boundaries for key regions such as the Great Attractor (Laniakea) and Coma superclusters. The study highlights the emergence of the Vela supercluster with the addition of new measurements, emphasizing the importance of continued observations in this region. 
    \item Within a distance of 150 $h^{-1}$Mpc, the measured  bulk flow  is 315 $\pm$ 40 km s$^{-1}$, consistent with previous measurements and predictions from the $\Lambda$CDM cosmological model. Beyond 150 \hmpc\ the study found that the bulk flow amplitude, shows increasing tension with $\Lambda$CDM, however this is largely a result of the dominance of the +SGY component. At present significant caution interpreting the large bulk flow at these scales should be exercised.  
\end{itemize}

Overall, the integration of new datasets into the CF4++ catalogue shows that the dynamical scale of homogeneity is not yet reached within the interval [200-300] $h^{-1}$Mpc from the observer, exhibiting a $2.28\sigma$ tension with $\Lambda$CDM, indicating that the local Universe still exhibits significant fluctuations in the distribution of mass and dynamics of galaxy motions. Future surveys and continued observations will further refine this cosmography, determining whether this tension is artificial due to current survey size limitations, as well as contributing to a more comprehensive understanding of the Universe's structure and evolution.

\section*{Data Availability}
The observational datasets used in this article are all publicly available from their original publications and also at \url{https://github.com/jrmould/fast}. The density and velocity grids and their associated uncertainties, as well as a sample code to obtain these values if given a position, assuming $B < 15$ and redshift $<$ 30,000 \kms, are available here \url{https://projets.ip2i.in2p3.fr//cosmicflows/}.

\begin{acknowledgements}
AD is supported by a KIAS Individual Grant PG 087201 at the Korea Institute for Advanced
Studies. HMC acknowledges support from the Institut Universitaire de France and from Centre National
d’Etudes Spatiales (CNES), France. JRM received support from the ARC CoE Centre for Dark Matter
Particle Physics (CDM, CE200100008). This research used services and data provided by the Astro Data Lab at NSF’s National Optical-Infrared Astronomy Research Laboratory.
Details are given in the github URL above.
\end{acknowledgements}

\bibliographystyle{aa} % style aa.bst
\bibliography{biblio}

\end{document}